# Monitoring subunit rotation in single FRET-labeled $F_oF_1$-ATP synthase in an anti-Brownian electrokinetic trap

Thomas Heitkamp[a], Hendrik Sielaff[a], Anja Korn[a], Marc Renz[a], Nawid Zarrabi[b], Michael Börsch[a,b,*]

[a] Single-Molecule Microscopy Group, Jena University Hospital, Friedrich Schiller University Jena, Nonnenplan 2 - 4, 07743 Jena, Germany
[b] 3rd Institute of Physics, University of Stuttgart, Pfaffenwaldring 57, 70550 Stuttgart, Germany

## ABSTRACT

$F_oF_1$-ATP synthase is the membrane protein catalyzing the synthesis of the 'biological energy currency' adenosine triphosphate (ATP). The enzyme uses internal subunit rotation for the mechanochemical conversion of a proton motive force to the chemical bond. We apply single-molecule Förster resonance energy transfer (FRET) to monitor subunit rotation in the two coupled motors $F_1$ and $F_o$. Therefore, enzymes have to be isolated from the plasma membranes of *Escherichia coli*, fluorescently labeled and reconstituted into 120-nm sized lipid vesicles to yield proteoliposomes. These freely diffusing proteoliposomes occasionally traverse the confocal detection volume resulting in a burst of photons. Conformational dynamics of the enzyme are identified by sequential changes of FRET efficiencies within a single photon burst. The observation times can be extended by capturing single proteoliposomes in an anti-Brownian electrokinetic trap (ABELtrap, invented by A. E. Cohen and W. E. Moerner). Here we describe the preparation procedures of $F_oF_1$-ATP synthase and simulate FRET efficiency trajectories for 'trapped' proteoliposomes. Hidden Markov Models are applied at signal-to-background limits for identifying the dwells and substeps of the rotary enzyme when running at low ATP concentrations, excited by low laser power, and confined by the ABELtrap.

## 1 INTRODUCTION

*Escherichia coli* F-type ATP synthase ($F_oF_1$-ATP synthase) is a membrane enzyme consisting of the transmembrane $F_o$ portion with subunit composition $ab_2c_{10}$, and the $F_1$ portion comprising $\alpha_3\beta_3\gamma\delta\varepsilon$ (Fig. 1A). The $F_o$ portion utilizes the electrochemical potential of protons over the membrane, that is the proton motive force, to power the synthesis of adenosine triphosphate (ATP) from adenosine diphosphate (ADP) and inorganic phosphate ($P_i$) in the $\alpha_3\beta_3$-subunits of the $F_1$ portion. Depending on the physiological conditions, the bacterial enzyme can also hydrolyze ATP to pump protons across the membrane in the other direction.

$F_oF_1$-ATP synthase is called a motor protein consisting of two nanomotors that work against each other[1, 2]. Both motors are hold together by a peripheral static connection of subunits. The rotary subunits of the motors are the ring of ten *c* subunits in $F_o$ and the subunits $\gamma$ and $\varepsilon$ in $F_1$, while the stator comprises subunits $ab_2\alpha_3\beta_3\delta$. In the $F_o$ motor, protons are transduced through a channel in subunit *a* to a negatively-charged residue in the *c*-subunit facing the membrane embedded *a*-subunit. After this residue is neutralized by the proton, electrostatic restrains force the *c*-ring to rotate forward. On the other end of the *c*-ring, a neutralized *c* subunit rotates from the hydrophobic lipid membrane environment towards the *a*-subunit interface, where it releases a proton through a second half-channel in subunit *a* to the opposite side of the membrane. The proton motive force provides the energy for the unidirectional rotation of the *c*-ring.

Subunits $\gamma$ and $\varepsilon$ in $F_1$ form a second motor that is elastically coupled to the *c*-ring[3, 4]. Subunit $\gamma$ forms the central rotary stalk that contacts the *c*-ring with its globular portion and intrudes the hexameric barrel of the three $\alpha\beta$-subunits of the stator. The asymmetric structure of subunit $\gamma$ contacts mainly one of the three $\alpha\beta$-pairs, thereby determining the opening and closing state of the three nucleotide binding sites in all three $\alpha\beta$-pairs.

---

* Email: michael.boersch@med.uni-jena.de or m.boersch@physik.uni-stuttgart.de ; http://www.m-boersch.org

During a 360° rotation of the γ subunit all three αβ-pairs cycle sequentially through the same conformational states, as proposed in the early 1980's[5] and suggested by the crystal structure of mitochondrial $F_1$-ATPase from bovine heart[6]. It was revealed in 1997 that the γ subunit of $F_1$ rotates in 120° steps during ATP hydrolysis[7], corresponding to the 3-fold symmetry of the three catalytic binding sites of the $α_3β_3$ barrel. Each rotational step is further divided into an 80° and a 40° substep[8, 9]. The former corresponds to the binding of an ATP molecule to an empty nucleotide binding site, whereas the later one is accompanied by the ATP hydrolysis reaction and product release. The rotational states correlate with the crystal structure[10].

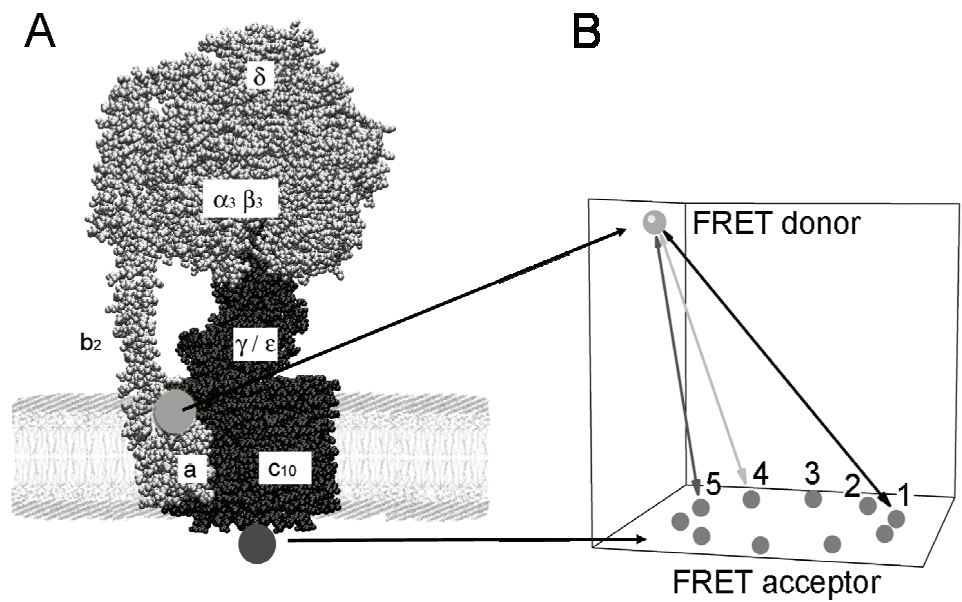

**Figure 1.** (**A**) Structure of the $F_oF_1$-ATP synthase from *Escherichia coli*. The rotary subunits γ, ε and $c_{10}$ are shown in black, the stator subunits are shown in grey. The membrane-embedded subunits *a*, $b_2$ and $c_{10}$ belong to the $F_o$ portion. For FRET-based monitoring of *c*-ring rotation two fluorophores were attached, one to the *a*-subunit (grey dot) and one to a single *c*-subunit (dark grey dot). (**B**) Expected distance changes during rotation of the *c*-subunit (labeled with FRET acceptor dye) with respect to the static position of the marker on the *a*-subunit (labeled with FRET donor dye). The symmetrically assigned stopping positions of the c-ring are named 1 for the longest and 5 for the shortest internal distances.

Our group studies single $F_oF_1$-ATP synthase molecules reconstituted in proteoliposomes to show the conformational changes and to resolve the stepsize of the rotor subunits in the holoenzyme *via* Förster resonance energy transfer (FRET) since 15 years[4, 11-39]. In a first set of experiments $F_oF_1$-ATP synthase molecules were labeled with two fluorescent dyes. For example, as shown in Fig. 1B one fluorescent dye EGFP was fused to stator subunit *a* and served as a FRET donor for the FRET acceptor dye Alexa568 that was coupled to one of the 10 rotor subunits $c^{[23]}$. During rotation the distance between the two fluorophores changed stepwise and sequentially. The shortest distance to the FRET donor was measured when the FRET acceptor dye on *c* had rotated to position 5 in Fig. 1B, and the longest distance was related to position 1. The *c*-ring rotated one position at a time yielding a sequence like 1→2→3→4→ as observed in a home-built confocal microscope for single-molecule FRET in solution. Rotation was driven either by ATP hydrolysis, or by a pH difference plus $K^+$ diffusion potential over the membrane to power ATP synthesis. In recent experiments we used a three color FRET approach to monitor the rotation of ε in $F_1$ and the *c*-ring in $F_o$ simultaneously[25, 38]. These experiments revealed the 10-step rotation of the *c*-ring, and correlated a single 120° step of subunit ε with three to four individual substeps of the *c*-ring.

The main limitation of our confocal single-molecule FRET approach using freely diffusing, liposome-reconstituted $F_oF_1$-ATP synthase is the short observation time. Owing to stochastic Brownian motion of the proteoliposomes through the confocal excitation and detection volume, the time trajectories have a maximum length in the range of several hundreds of milliseconds. The average diffusion time of our proteoliposomes through a detection volume of several femtoliters is about 30 ms. Therefore we need to find alternative methods to keep the FRET-labeled enzyme in solution within the confocal volume.

The so-called <u>A</u>nti-<u>B</u>rownian <u>E</u>lectrokinetic <u>T</u>rap (ABELtrap) was presented in 2005 by A. E. Cohen and W. E. Moerner. Their microfluidic device can hold single small particles like fluorescent beads, liposomes, DNA or proteins in solution[40-42]. The ABELtrap comprises a 1-µm shallow trap region with four access channels, and one platinum electrode is placed in each channel. Fluorescence of the particle is used to localize its position in x and y coordinates. Voltages are supplied to the electrodes to move the particle by electrophoretic and electroosmotic forces. Thereby the arbitrary Brownian motion of the particle is compensated and the particle remains stationary. Several different ABELtrap versions have been realized[43-47], with very fast feedback times in the microsecond time range[48-52].

Recently we have built an ABELtrap using a confocal laser pattern which is controlled by a field-programmable gate array (FPGA). This ABELtrap could hold 20-nm fluorescent polystyrene beads in solution for more than 8 seconds, that is with a 1000-fold prolongated observation time (N. Zarrabi et al, Proc. SPIE 8587 (2013), *in press*). Here, we describe the current status of $F_oF_1$-ATP synthase preparations and simulate the recovery of FRET levels by Hidden-Markov-Models (HMMs) at the lower limit of signal-to-background ratios, as expected for maximum observation times of the enzyme in the ABELtrap using low laser excitation power for FRET measurements.

## 2 EXPERIMENTAL PROCEDURES

### 2.1 Preparation and characterization of $F_oF_1$-ATP synthase

The $F_oF_1$-ATP synthase is prepared from the plasma membranes of *Escherichia coli* bacteria. The details of our current preparation procedures are given below.

*Escherichia coli strains and growth conditions*

For the expression of the *E. coli* atp genes coding for the $F_oF_1$-ATP synthase, strain RA1 *(F⁻ thi rpsL gal Δ(cyoABCDE)456::KAN Δ(uncB-uncC) ilv:Tn10)*[53] was used which lacks a functional $F_oF_1$-ATP synthase. This strain was transformed with the plasmid pRA100 carrying the structural genes of the atp operon under control of the atp-promoter[54]. The cells were grown in a modified complex medium (0.5 g/l yeast extract, 1 g/l tryptone, 17 mM NaCl, 10 mM glucose, 107 mM $KH_2PO_4$, 71 mM KOH, 15 mM $(NH_4)_2SO_4$, 4 µM uracil, 50 µM $H_3BO_3$, 1 µM $CoCl_2$, 1µM $MnCl_2$, 2 µM $ZnCl_2$, 10 µM $CaCl_2$, 3 µM $FeCl_2$, 0.5 mM $MgSO_4$, 0.5 mM arginine, 0.5 mM isoleucine, 0.7 mM valine, 4 µM thiamine and 0.4 µM 2,3-dihydroxybenzoic acid). Cells were grown in a 10 L fermenter (FerMac 320, Electrolab, UK) at 37°C at 800 rpm, with a fresh air supply at 4.5 L/min, to the late logarithmic phase. Cells were subsequently harvested by centrifugation in a Sorvall Evolution RC centrifuge (Thermo Fisher Scientific, USA) at 10.000 × g and 4° C for 5 min. Cells were washed with 50 mM MOPS/KOH pH 7.0, 150 mM KCl, 10 mM $MgCl_2$, 0.2 mM EGTA, 0.2 mM DTT and 0.1 mM PMSF, and pelleted again at 10.000 × g for 5 min.

*Purification of the $F_oF_1$-ATP synthase*

*E. coli* $F_oF_1$-ATP synthase without Histidin-tags was purified as already described[55] with several modifications. All subsequent steps were done at 4° C. Cells were resuspended in 20 mM Tris-HCl pH 8.0, 140 mM KCl, 5 mM $MgCl_2$, 5 mM para(6)-aminobenzoic acid (PABA), 2 mM DTT, 10% (v/v) glycerol, 0.002% (w/v) PMSF and a spatula point DNase I, and lysed by one passage through a cell homogenizer (PandaPlus 2000, GEA Niro Soavi, Italy) at 1200 bar. Cells were cooled before and after lysis. Cell debris was pelleted by two centrifugations at 25000 × g for 20 min in a cooled centrifuge (Sorvall Evolution RC, Thermo Fisher Scientific, USA). Membranes were subsequently harvested by centrifugation at 185,000 × g for 1.5 h in an ultracentrifuge (Optima XP, Beckman Coulter, USA). The membrane pellet was resuspended in 50 mM Tris-HCl pH 8.0, 100 mM KCl, 5 mM $MgCl_2$, 5 mM PABA, 2% (w/v) sucrose, 10% (v/v) glycerol and 0.001% (w/v) PMSF and centrifuged again as described above (Optima XP, Beckman Coulter, USA). The washed membranes were resuspended in 20 mM MES/Tricine-KOH pH 7.0, 5 mM $MgCl_2$, 2 mM DTT and 0.001% (w/v) PMSF to a final volume of 10 ml buffer per g membranes, and membrane proteins were solubilized using 1.75% (w/v) DDM (Glycon, Germany) for 30 min on ice with gentle stirring. Solubilized proteins were separated by centrifugation at 220,000 × g for 90 min, and subsequently precipitated by a two-step $(NH_4)_2SO_4$ precipitation. In the first step, impurities were precipitated with 45% (v/v) saturated $(NH_4)_2SO_4$ and pelleted at $25000 \times g$ for 15 min. Then, $F_oF_1$ was precipitated with 65% (v/v) saturated $(NH_4)_2SO_4$ and pelleted the same way.

The precipitated protein was resolved in 2 ml 40 mM MOPS-KOH pH 7.5, 80 mM KCl, 4 mM $MgCl_2$, 2 mM DTT, 2% (w/v) sucrose, 10% (v/v) glycerol, 1% (w/v) DDM and 0.001% (w/v) PMSF, loaded on a self-packed XK16/100 Sephacryl S300 size exclusion column (GE-Healthcare, USA), which was externally cooled with a chiller FL601 (Julabo, Germany) and was equilibrated with the same buffer containing 0.1% (w/v) DDM. Proteins were eluted using an Äkta PrimePlus FPLC system (GE-Healthcare, USA) at a flow rate of 0.6 ml/min. Subsequently the peak fractions of the enzyme were loaded separately on a Poros HQ 20 (4.6 x 100 mm) ion exchange column (Applied Biosystems, USA) equilibrated with the size exclusion buffer listed above and connected to a modular FPLC system (LCC500 controller, P500 pumps, Frac100 fractionator, UVM II UV monitor; Pharmacia, Sweden). The column was washed with the same buffer for 5 column volumes, and the proteins were eluted by a KCL gradient over 20 column volumes up to 1.5 M KCl.

The $F_oF_1$-containing ion exchange fractions were pooled, concentrated by precipitating the protein with 65% (v/v) saturated $(NH_4)_2SO_4$ and resolved in 0.5 ml MOPS-DDM buffer as described above. The proteins were further purified by a second size exclusion column using the same buffers and specifications as above except that now a self-packed XK16/100 Sephacryl S400 size exclusion column (GE-Healthcare, USA) was used. The peak fractions of the enzyme were directly tested for ATP hydrolysis activity and a small fraction of the enzyme was reconstituted in liposomes for subsequent ATP synthesis measurements. The remaining $F_oF_1$ solution was shock-frozen in liquid nitrogen in 500 µl aliquots (cryo straws) and stored at -80 °C.

*ATP hydrolysis measurements*

ATP hydrolysis rates were measured on the basis of established protocols[56]. Briefly, 2 µl of the Sephacryl S400 peak fractions containing 1-2 µg of $F_oF_1$ were added to 0.5 ml 50 mM Tris-Acetic acid pH 8.5, 10 mM ATP and 4 mM Mg-acetate and incubated at 30° C for 5 minutes. The ATP hydrolysis reaction was stopped by adding 0.5 ml of 10% (w/v) SDS, and the free phosphate was estimated by measuring the absorbance at 700 nm after adding of 0.5 ml ferrous ammonium sulfate-$H_2SO_4$ ammonium molybdate reagent[57]. The specific LDAO activation of ATP hydrolysis (ATPase activity) was determined by adding 0.5% (v/v) LDAO to the reaction mixture.

*Reconstitution of $F_oF_1$*

Reconstitution into proteoliposomes was carried out according to[58]. Briefly, 500 µl of preformed liposomes (prepared according to[59]) were mixed with 2.5 µl 1M $MgCl_2$ and 40 µg of purified protein and then diluted with 20 mM Tricine-NaOH pH 8.0, 20 mM succinate, 80 mM NaCl and 0.6 mM KCl to yield a volume of 920 µl. 80 µl 10% (v/v) Triton X-100 was added to destabilize the membranes. After 1 h incubation with gentle stirring, 320 mg of pretreated BioBeads SM-2 (Biorad, USA) were added to remove the detergent. After one additional hour, the proteoliposomes were separated from the BioBeads and immediately used for ATP synthesis measurements. These proteoliposomes had a final $F_oF_1$ concentration of 80 nM.

*ATP synthesis measurements*

The rate of ATP synthesis was measured at 21 °C as described earlier[59]. 20 µl of proteoliposomes were incubated for 3 min with 80 µl 20 mM succinate, 5 mM $NaH_2PO_4$, 0.6 mM KOH, 2.5 mM $MgCl_2$, 0.1 mM ADP and 20 µM valinomycin titrated to pH 4.5. The ATP synthesis reaction was started by adding 225 µl 200 mM Tricine, 5 mM $NaH_2PO_4$, 160 mM KOH, 2.5 mM $MgCl_2$ and 0.1 mM ADP titrated to pH 8.8 premixed with 75 µl Bioluminescence Assay Kit CLS II (Roche, Switzerland) to 15µl of pre-incubated liposomes in a luminometer (Sirius luminometer, Berthold Detection Systems, Germany). In this setup, the synthesized ATP was measured directly in the reaction mixture during the ongoing reaction for 30 s. The injection was done automatically in the luminometer. The emitted luminescence was measured every 0.2 s, and the data were transferred to a computer for analysis.

*Other methods*

Protein concentrations were determined using the modified Lowry method[60]. SDS-polyacrylamide gels were made as described[61] with acrylamide concentrations between 12-14%, and silver-stained according to[62].

## 2.2 Principle of the ABELtrap

The microfluidic device keeping the liposome-reconstituted $F_oF_1$-ATP synthase in solution within the laser focus is shown in Fig. 2. Diffusion of the proteoliposome is limited to the x- and y-directions within the 1-µm shallow trapping region which is located between the cover glass and the PDMS sample chamber. Electrokinetic forces comprise electrophoretic and electroosmotic forces to counteract the actual movement of the proteoliposome (Fig. 2A). Potentials are applied to the four platinum electrodes to push back the proteoliposome (Fig. 2B).

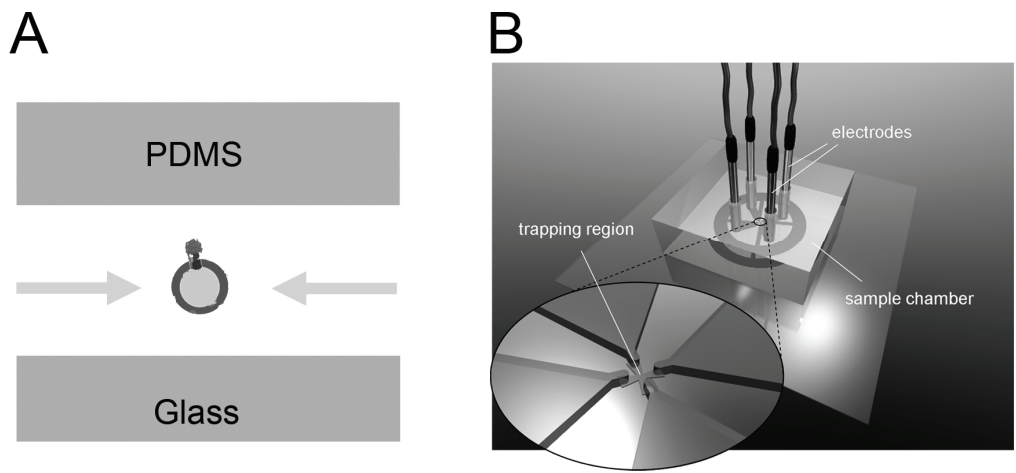

**Figure 2.** (**A**) Schematic of the confinement in the ABELtrap. Electrokinetic forces keep the proteoliposome in place. (**B**) PDMS sample chamber with flat trapping region and platinum electrodes[32].

In our confocal ABELtrap setup, the laser focus (491 nm, Cobolt Calypso, Sweden) is moved within a 2 µm × 2 µm area inside the 1-µm shallow trapping region. Once a fluorescent particle, that is the FRET-labeled $F_oF_1$-ATP synthase, is excited, the position of maximum fluorescence intensity is estimated and a feedback voltage is created to move the particle into the center of the laser pattern. We use a pair of AOBDs (Gooch & Housego, NEOS technologies, 46080-3-LTD) to shift the laser focus in x and y coordinates. Two single photon counting APDs detect FRET donor and FRET acceptor photons separately. Their signals are used by the field-programmable gate array (FPGA) as the feedback to control the voltages at the electrodes[49]. The FPGA also generates the pattern of laser positions.

## 2.3 Simulation of FRET data mimicking *c*-subunit rotation in $F_oF_1$-ATP synthase

To evaluate the minimum signal-to-background ratio required for subsequent FRET data analysis by Hidden Markov Models (HMMs), we simulated the stepwise rotation of the *c*-ring of $F_oF_1$ using a Monte-Carlo simulation. Therefore, a single FRET-labeled particle is placed into a virtual box. Within this box, a three-dimensional ellipsoid is centered representing the three-dimensional detection volume of a confocal experiment. Previously, the FRET-labeled particle could diffuse freely in and out of the ellipsoid[20]. Here, we changed the simulations so that the FRET-labeled particle was trapped inside the ellipsoid as soon as it diffused to the boundaries of the ellipsoid. Thereby the confinement achieved by the ABELtrap experiment was simulated. Depending on the given photophysical rates for photobleaching of the FRET donor and FRET acceptor dyes, the particle was forced to stay in the ellipsoid until the signal disappeared irreversibly. Afterwards, a new freely-diffusing FRET-labeled particle was generated in the box but outside of the ellipsoid, and the ABELtrap simulations proceeded. A high background count rate on both detection channels and low photon count rates with shot-noise-limited intensity fluctuations were applied yielding nearly constant sum intensities of both FRET donor and acceptor photon counts.

## 2.4 Hidden Markov Model-based FRET level analysis

The generated FRET time trajectories of the simulated particles in the ABELtrap were analyzed using previously described Hidden Markov Models with the given number of 10 states. These 10 HMM states corresponded to 5 distinguishable FRET levels for symmetry reasons of the *c*-ring of $F_oF_1$-ATP synthase. In contrast to our previous HMM approach to find FRET levels in freely-diffusing proteoliposomes[20, 23] we used a Gaussian distribution for each FRET level with variable widths. After the first round of HMM learning of the 5 FRET levels and dwell times, the resulting FRET levels and corresponding dwell times were used to assign these FRET levels to the FRET time trajectories. The dwell time histograms of the assigned FRET levels were fitted with monoexponential decay functions to unravel the deviations between the learned HMM states and the assigned FRET levels.

## 3 RESULTS

### 3.1 Preparation of a fully active $F_oF_1$-ATP synthase

We established the isolation procedures of a fully active $F_oF_1$-ATP synthase from *E. coli* as published by Peter Gräber's laboratory at the University of Freiburg[14, 16, 59]. Our changes included the use of a new 10 L fermenter system for cell growth, a different type of cooled cell homogenizer to prepare the plasma membranes, and two distinct FPLC systems to run the size exclusion and ion exchanges columns. The holoenzyme $F_oF_1$-ATP synthase could be purified to a high purity using a three column purification protocol. In the first purification step a size exclusion column (Fig. 3A left lane, 'S300', Sephacryl S300) was used for desalting the DDM-dissolved membranes. Furthermore, this step allowed separating $F_oF_1$ from a protein impurity, which could not be separated by the subsequent ion exchange purification step. The corresponding SDS-PAGE of the proteins showed, that the protein sample still contained a lot of impurities. However, the majority of these impurities could be separated in the next step by the Poros HQ 20 ion exchange column (Fig. 3A, middle lane, 'IEC'). $F_oF_1$ bound most probably *via* the $F_1$ portion to the ion exchange column and eluted at a chloride concentration of about 300 mM, whereas the very tightly bound contaminants eluted at about 900 mM chloride concentration. The application of a final size exclusion column (Fig. 3A, right lane, 'S400'; Sephacryl S400) increased the purity of the $F_oF_1$-ATP synthase preparation further, and was used to exchange the high salt concentration buffer.

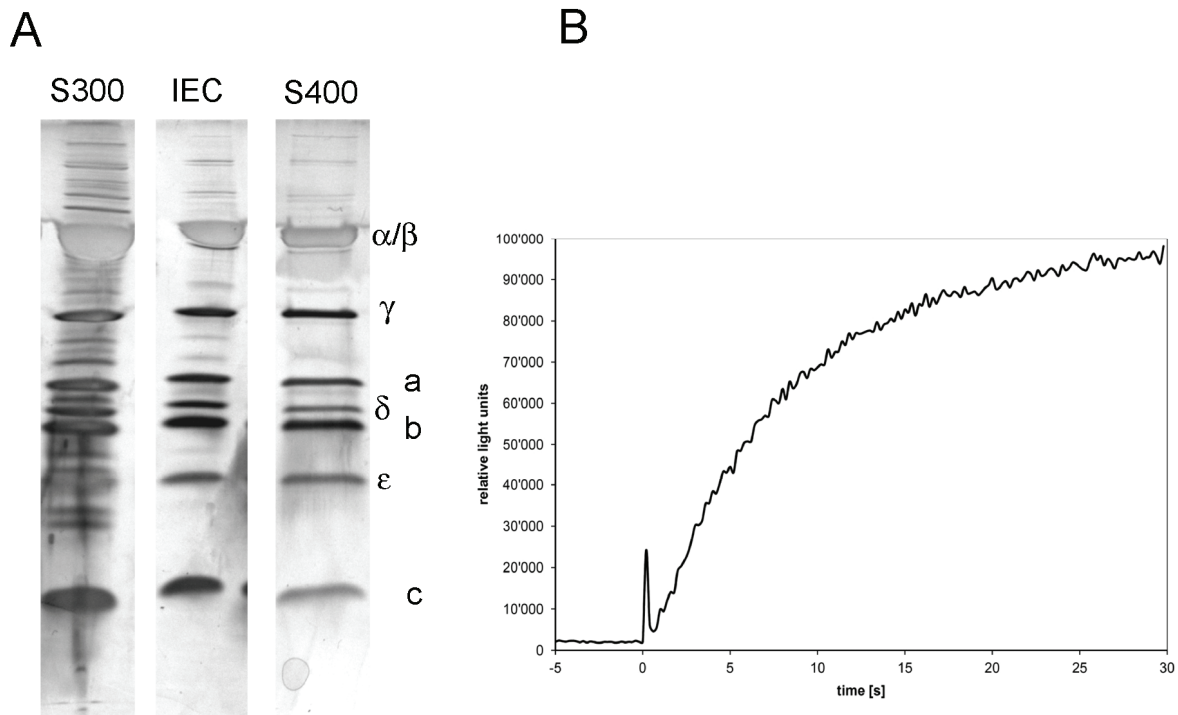

**Figure 3.** (**A**) Purification of $F_oF_1$-ATP synthase as revealed by silver-stained 12% SDS-PAGE of the major purification steps. After the first size exclusion column (Sephacryl 'S300') the sample still contained a lot of protein impurities which were effectively removed from $F_oF_1$ by the ion exchange chromatography ('IEC', Poros HQ 20). The final size exclusion column (Sephacryl 'S400') yielded a pure enzyme, with all subunits from $F_1$ (α, β, γ, δ, ε) and $F_o$ (*a*, *b*, *c*) present. (**B**) Measurement of ATP synthesis rates using bioluminescence of the luciferin / luciferase reaction. Injection of the basic buffer with luciferin and luciferase to the proteoliposomes resulted in a small peak at time '0' s. Rising luminescence intensities indicate ATP synthesis.

The ATP hydrolysis rate (ATP turnover) of the soluble $F_oF_1$ at 30° C was around 180 s$^{-1}$, which is in the same range as published activities[63]. Furthermore, ATP hydrolysis could be specifically activated 3.3-fold by the addition of LDAO. The liposome-reconstituted $F_oF_1$ complex was tested for ATP synthesis activity using the luciferin / luciferase bioluminescence reaction. Equilibration of proteoliposomes in acidic buffer increased the proton concentration inside the liposome. An initial ΔpH > 4 and an electric potential across the liposome membrane were created to drive ATP synthesis by injection of a basic buffer to the proteoliposome suspension. The immediate rise of relative luminescence light units indicated the onset of ATP synthesis (Fig. 3B) The preliminary ATP synthesis rates were qualitatively comparable to the literature values[58].

### 3.2 Simulating *c*-ring rotation of $F_oF_1$-ATP synthase in the ABELtrap

Monte-Carlo (MC) simulations were applied to evaluate the possible observation time increase by trapping $F_oF_1$-ATP synthases in solution, and using low laser excitation power to avoid fast photobleaching of the FRET fluorophores. The box size for a freely diffusing single FRET-labeled particle was set to 2.6 µm × 2.6 µm (x and y direction) and 13.2 µm (z direction) yielding a volume of 89.2 fl. Within this box a three-dimensional Gaussian ellipsoid was placed in the center. The boundaries were $\sigma^2$=0.65 µm for the x- and y radii and $\sigma^2$=3.3 µm for the z axis resulting in a 'confocal' volume of 5.8 fl. The diffusion coefficient of the FRET-labeled particle was set to D=3*10$^{-8}$cm$^2$/s, i.e. 100-fold smaller than for a single fluorophore in water. Once the particle hit the boundaries of the ellipsoid, the particle was considered "trapped", i.e. could not escape from the ellipsoid anymore.

Simulated photon count rates of the FRET-labeled particle were 20 kHz (sum of both fluorophores). The photobleaching probability of the FRET donor fluorophore was simulated given a mean of 30.000 emitted photons before bleaching. The FRET acceptor was considered to be infinitely photostable in these simulations. High background count rates of 10 kHz for both detection channels were chosen to mimic the experimentally detected luminescence from a glass cover slide and PDMS which we used for our ABELtrap sample chambers. Ten time trajectories with 60 s each were simulated, with 215 photon bursts in total.

The rotation of the *c*-ring in $F_oF_1$-ATP synthase proceeds stepwise in 10 steps for 360° rotation. For symmetry reasons we assigned 5 distinct FRET efficiencies to these 10 steps (see Fig. 1B). The sequence of FRET levels with FRET efficiencies from 0.3 to 0.7 was set as follows: simulated state 1 at FRET efficiency 0.3 , state 2 at 0.4, state 3 at 0.5, state 4 at 0.6, state 5 at 0.7, state 6 at 0.7, state 7 at 0.6, state 8 at 0.5, state 9 at 0.4, and state 10 at 0.3. A directionality of rotation was implemented using a likelihood ratio of 20:1 for the forward stepping in the state sequence 1→2→...→10→1→..., i.e. back stepping was allowed but rare. Dwell times for each of the 10 states were set to 20 ms. The time resolution in the MC simulations were 100 µs per data point. Time traces were subsequently binned to 1 ms. Our simulated data file format was similar to the experimental TCSPC data format from Becker&Hickl, so that we could use the software 'Burst_Analyzer' (Becker&Hickl, Berlin, Germany) for data visualization and threshold-based photon burst identification.

FRET time trajectories were analyzed using a 10-state HMM with 5 different FRET efficiencies. The HMM did not use assumptions of a preferred direction of rotation. A constant background of 10 counts per ms (10 kHz) on both channels was subtracted from the simulated FRET donor and acceptor intensity data to get the correct FRET efficiencies. Starting values for the HMM were set to the five FRET efficiencies 0.1667, 0.3333, 0.5000, 0.6667 and 0.8333, with equal small variances of 0.02. Starting dwell times were set to 50 ms.

Iterative learning yielded the HMM results with FRET efficiencies and variances for the 5 FRET levels and the associated dwell times: FRET level 1 at 0.3064 with variance 0.0077 and dwell time 19.6 ms; FRET level 2 at 0.3961 with variance 0.0115 and dwell time 13.3 ms; FRET level 3 at 0.5023 with variance 0.012 and dwell time 13.4 ms; FRET level 4 at 0.6081 with variance 0.0113 and dwell time 12.8 ms; and FRET level 5 at 0.6983 with variance 0.0077 and dwell time 18.6 ms. The recovery of FRET levels was nearly perfect. The dwell times were estimated slightly too short, but the longer dwell times for FRET levels 1 and 5 correctly identified the fact that these FRET levels have the doubled dwell time due to hidden transitions within the same FRET level, because state 1 and 10 have the same FRET efficiency (as well as states 5 and 6) and state transitions 10→1 and 5→6 will result in twice as long dwells for the corresponding FRET levels 1 and 5.

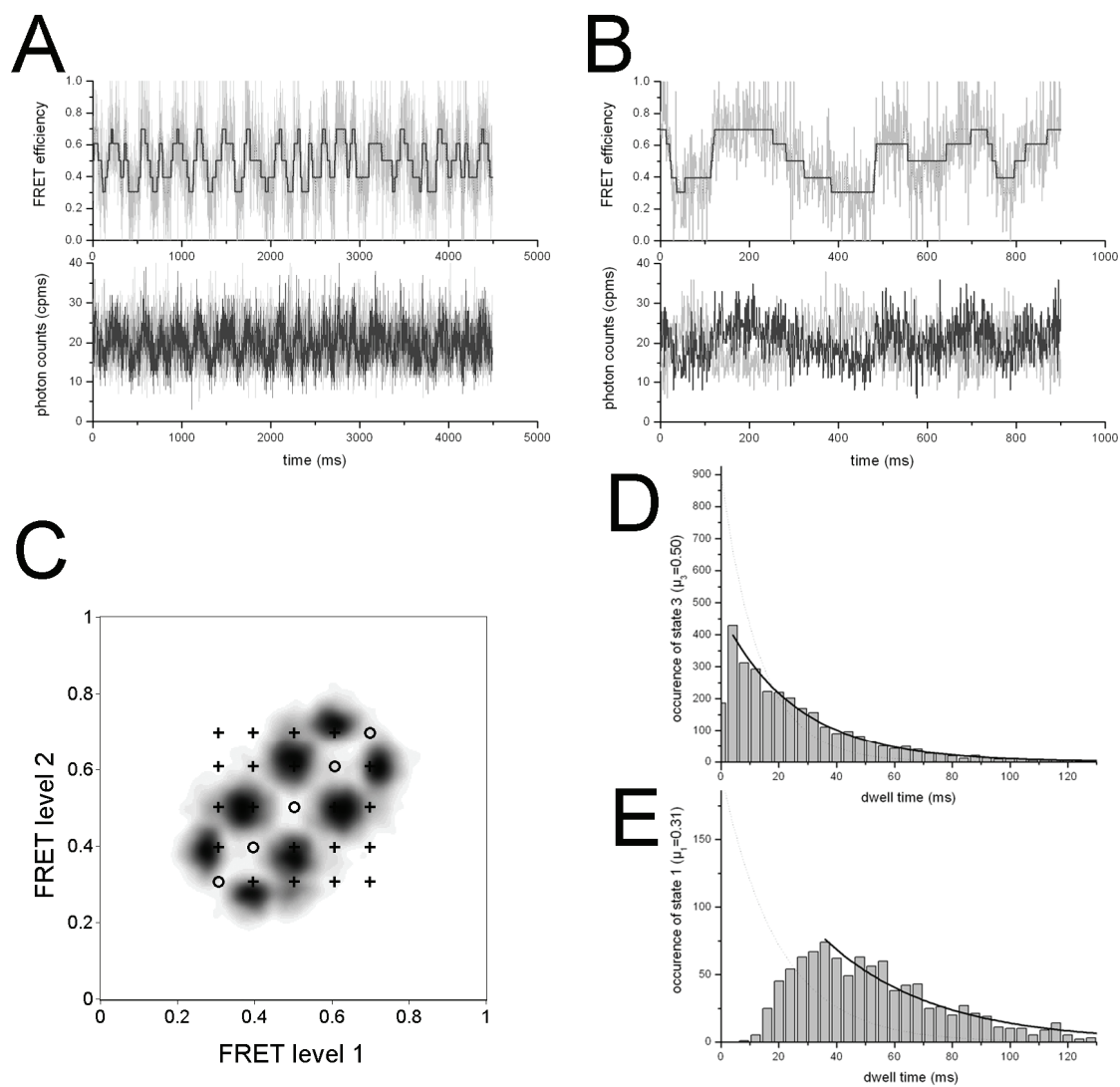

**Figure 4.** Simulated FRET data and FRET level recovery by HMM (see text for detailed description). (**A**), (**B**) Time trajectories of two photon bursts with simulated FRET levels (dotted line) and assigned FRET levels (black stepped line) in the upper panels. (**C**) FRET transition density plot of assigned FRET levels with simulated transitions (black crosses). (**D**), (**E**) Dwell time histograms and fits for assigned FRET levels 1 at 0.31 and 3 at 0.5. Black lines are monoexponentional fits to the maxima of the histograms, dotted lines are expected from the simulations.

After learning of FRET levels and dwell times using the 10-state symmetric HMM, the assignment of FRET levels in the simulated time trajectories was achieved. 9326 FRET levels were assigned in 215 photon bursts. The results of the HMM assignments are summarized in Fig. 4. Two examples of photon bursts in Fig. 4A and 4B show many transitions of FRET levels. The lower panels contain intensity trajectories of FRET donor (light grey) and acceptor (dark grey). The upper panels show the FRET efficiencies per data point (light grey), the simulated FRET efficiency trajectories (dotted lines) and the HMM-assigned FRET levels (black lines with steps). Manual comparison revealed that most of the FRET levels were recovered correctly. However, short dwells were missed more often, especially for the lowest and highest FRET efficiencies (see Fig. 4A, between time 2400 ms and 3300 ms; these FRET levels were also overlooked and lost in Fig. 4B at times between 500 and 700 ms).

The FRET transition density plot[64] in Fig. 4C showed 8 major transitions, i.e. from FRET efficiency 0.7 to 0.6, from 0.6 to 0.5, and so on. Black crosses represented the simulated expected transitions with changing FRET levels, and black rings corresponded to transitions to the same FRET efficiency (i.e. transitions between HMM states 1↔10 and 5↔6). Only small populations of transitions between non-neighboring FRET levels were assigned (for example, transitions between 0.5 and 0.3), but hardly any larger FRET level transitions. The largest deviations from the given FRET level transitions were found for the two neighboring low FRET (0.3↔0.4) and high FRET (0.6↔0.7) transitions.

The dwell times associated with the assigned FRET levels were fitted by monoexponential decay functions from the maximum of each distribution. In general the assigned dwell times were significantly longer than those learned by the HMM. The dwell time of assigned FRET level 0.5 was 26 ms, which is in reasonable agreement with the simulated dwell time of 20 ms, but in contrast to the HMM dwell time of 13.4 ms (Fig. 4D). The dwell time of assigned FRET level 0.31 was 37.9 ms (Fig. 4E). This was in good agreement with the expected dwell time of 40 ms for this FRET level, i.e. twice the dwell time of state 1 (due to the transition between state 10 to 1 at the same FRET level 0.3). However, dwell times of assigned FRET levels 0.40 and 0.61 were also significantly longer (38.6 ms or 37.8 ms, respectively) than the simulated 20 ms for these states. The dwell time histograms clearly deviated from a distribution with single monoexponential decay. Due to missing short dwell times additional rising components for fitting of the distributions were needed.

## 4 DISCUSSION

According to the provided biochemical data we have re-established successfully the preparation of $F_oF_1$-ATP synthase from *E. coli*. The procedures yielded active enzymes for both ATP hydrolysis and ATP synthesis. The distinct components used for the purification steps allowed to obtain the enzyme from *E. coli* in high purity for the subsequent single-molecule FRET studies of conformational dynamics, like rotary motor movements in $F_o$ or $F_1$, or regulatory changes in ε.

To overcome the observation time limitations of our previous single-molecule FRET experiments with freely-diffusing reconstituted enzymes, we had built an ABELtrap at the University of Stuttgart using an EMCCD camera for particle localization[2, 65]. Now we have re-built a new version of a faster ABELtrap that can hold fluorescent nanobeads in solution for more than 8 seconds. Because the nanobeads had a mean diameter of 20 and the proteoliposomes have a diameter of 100 to 200 nm, we expect that trapping FRET-labeled, reconstituted $F_oF_1$-ATP synthases can be accomplished.

Monitoring proton-driven *c*-ring rotation on single FRET-labeled $F_oF_1$-ATP synthases during ATP synthesis was achieved using a maximum proton motive force[23] which dissipated within few minutes. ATP hydrolysis-driven *c*-ring rotation was observed only at high [ATP] resulting in a mixture of small and large step sizes. Refined FRET measurements at lower [ATP] were needed, but could not be measured so far with the freely diffusing enzymes. Holding single FRET-labeled enzymes in the ABELtrap would allow these experiments. Prolonged observation times for single FRET-labeled $F_oF_1$-ATP synthases require a reduced laser excitation power to avoid premature photobleaching of the dye molecules. We found that the ABELtrap exhibited higher background photon count rates owing to the laser focus placed across a shallow trapping region and hitting both cover glass and the PDMS material of the sample chamber. Therefore we simulated FRET data at low photon count rates and in the presence of a high background.

FRET time trajectories based on low photon count rates of 20 kHz for FRET donor plus acceptor and in the presence of 10 kHz background on both channels were analyzed using 10-state Hidden Markov Model. Rotation of the *c*-ring was expected to result in 5 distinguishable FRET levels because of ring symmetry. The HMM analysis recovered these 5 simulated FRET levels based on 9326 assigned levels, given a pre-defined model of 10 HMM states with twofold degeneracy. The learned dwell times were estimated slightly too short. However, when FRET levels had been assigned within all time trajectories based on the HMM-learned FRET levels and transition likelihoods, the respective dwell time distributions of assigned FRET levels were found too long. We noticed in the assigned FRET levels that short dwells at the lowest and the highest FRET efficiencies were often missed. This explained why the dwell time distributions of the two lowest and the two highest FRET efficiencies could not be fitted by a monoexponential decay function but required an additional rising component.

In summary, successful recovery of FRET levels by HMM was possible at the low signal-to background ratio using a data set containing many FRET levels. High background appeared less critical for the HMM than small signal and short dwell times. As a consequence we plan future single-molecule FRET experiments for conditions with increased photon count rates at the expense of slightly faster photobleaching and shortened observation times. Then, the stepsize of $c$-ring rotation in $F_oF_1$-ATP synthase will be unraveled for a broad variety of rotor speeds using single-molecule FRET-based analysis of enzymes confined in solution by the ABELtrap.

**Acknowledgements**

This work was supported in part by the DFG grants BO 1891/10-2 and BO 1891/15-1 to M.B.. Financial support by the Baden-Württemberg Stiftung (by contract research project P-LS-Meth/6 in the program "Methods for Life Sciences") to built an ABELtrap is gratefully acknowledged. The authors want to thank Dr. Andrea Zappe (3[rd] Institute of Physics, University of Stuttgart, Germany), Prof. Dr. T. Duncan (SUNY Upstate Medical University, Syracuse, NY, USA) and Prof. Dr. P. Gräber (University of Freiburg, Germany) for their ongoing support in $F_oF_1$-ATP synthase mutants and enzyme preparations.